# Quantum Criticality and the $\alpha/\delta$ Puzzle

G. Chapline

Lawrence Livermore National Laboratory, Livermore, CA 94551

**Abstract**

In an overview of the elemental actinides Np and Pu stand out because of their anomalously low melting temperatures and the variety of complex phase transitions that occur in these elements and their alloys as a result of relatively modest changes in temperature and pressure. In this paper we suggest a novel explanation based on an analogy between the evolution of the actinide ground state as a function of spin-orbit coupling and the behaviour of thin film superconductors in a magnetic field. The key point is that in "bad metals" spin-orbit interactions give rise to low energy monopole-like solitons with quantized spin currents, which play much the same role as Abrikosov vortices in thin film superconductors. In Np and $\alpha$-Pu these solitons form an ordered solid, while in impurity stabilized $\delta$-Pu they form a pair condensate. This provides a simple explanation for the heretofore unexplained phenomenology of $\alpha/\delta$ transition. Near room temperature $\delta$-Pu represents a novel form of condensed matter: a "Planckian metal" analogous to the quark-gluon plasma.

At its onset the Manhattan Project was faced with the problem that the lattice structure and density of elemental Pu seemed to depend on how it was prepared. Despite the passage of 65 years an explanation of this puzzle has proved elusive. It has been suggested [1] that the underlying reason for this behaviour is that the ground state of elemental Pu lies near to a quantum critical point (QCP) of some kind. Indeed, the negative thermal expansion of pure $\delta$-Pu is by itself suggestive of critical behaviour analogous to the Neel point of an antiferromagnet. Of course pointing out that as a function of atomic number the actinide ground state may have a quantum critical point doesn't explain why this happens. For a long time the conventional view has been that the unusual properties of elemental Pu and the $\alpha/\delta$ transition in particular are somehow a consequence of a transition between itinerant $f$-electrons in the lower actinides and localized $f$-electrons in the higher actinides [2]. However it is now reasonably clear that the $f$-electrons in the pure elements do not become localized on an atomic scale as one moves from Np to Am. Indeed, studies of Pu/Am alloys [3] suggest that the $f$-electron bands do not change in any dramatic way as a function of Am concentration. On the other hand, as Harrison [4] has emphasized, the difference in the volume and free energy of the $\alpha$ and $\delta$ phases suggests that the $f$-electrons in the $\delta$ phase do not contribute to metallic binding. In the following we shall argue that this quasi-localization is very similar to the behaviour of electrons in disordered Type II thin film superconductors when the strength of an applied magnetic field exceeds a critical value, and the electrons form a "boson glass" [5].

Actually there are hints that the unusual behaviour of Np/Pu may also be related to the physics of high temperature superconductors with no external field. For example, a plot showing the superconducting and magnetic ordering temperatures of the elemental actinides as a function of atomic number is reminiscent of the phase diagram for the hole-doped superconducting cuprates as a function of doping [6]. In both cases there is region of the phase diagram where these materials are neither superconducting nor anti-ferromagnetic, and the electronic properties are anomalous. The striking correspondence between the Np/Pu/Am region of the actinide phase diagram and the "psuedogap" region of the high $T_c$ phase diagram between bulk superconductivity and antiferromagnetism may well be a hint that the underlying physics in the two cases is the same. The connecting link between the two cases, which could lead to deciphering quantum criticality in both the elemental actinides and high $T_c$ superconductors, is that in both situations spin-orbit interactions transform a gossamer density of paired itinerant electrons [7] into a ground state containing a condensate of topological solitons.

Just as an external magnetic field acts as a tuning parameter for the ground state of thin film superconductors, so it is actually quite plausible that the strength of spin-orbit interactions for the *f*-electrons in the actinides acts as a tuning parameter for the nature of the electronic ground state of the elemental actinides. Indeed spin orbit interactions in the actinides can be thought of as an effective magnetic field – the main difference being that the effective magnetic field depends on the electron momentum. In addition, it has recently been shown that the importance of spin-orbit interactions varies in a systematic way as one moves across the actinides [8]. In particular, measurements of the branching ratios for 4d → $5f_{5/2}$ and 4d → $5f_{7/2}$ transitions in the actinides using TEM electron energy loss spectroscopy suggest that spin-orbit effects increase dramatically in importance as one moves from U to Pu and Am. *Our proposal for understanding the unusual behaviour of the elemental actinides in the vicinity of Np/Pu/Am rests on the idea that as the strength of spin-orbit interactions for 5f electrons in these elements increases it reaches a critical value in the vicinity of Pu analogous to the critical strength of an external magnetic field where a disordered Type II thin film superconductor undergoes a superconducting-insulating transition.* (see Table 1). The picture that emerges for the α/δ transition is that the α and impurity stabilized δ phases of Pu are nearly degenerate because they are near to a quantum critical phase transition where the tuning parameter is either the strength of the spin-orbit interaction or disorder (see Fig.1). The actual transition is first order and involves a change in lattice structure due to a change in phonon entropy, but it is known from ab initio calculations [9] that the internal energies of the α and δ phases are nearly the same. This picture provides a simple explanation for the phenomenology of the α/δ transition. In particular, this explains why δ-Pu is stabilized by a wide variety of impurities and α-Pu is stabilized by U. Evidently in the case of δ-Pu it is not the chemical nature of the impurity that is important, but the disorder it creates.

The physical effects of spin-orbit interactions in a solid are quite different in good metals and metals with a low carrier density. It turns out that a mathematical basis for understanding spin-orbit effects in layered metals with a low carrier density was developed some years ago by the author [10]. This work generalized a tight binding theory of the "parity anomaly" in graphene that had been proposed a few years earlier by Haldane [11]. Although at the time neither of us realized that our mathematics might be relevant to understanding spin-orbit effects in metals with a low carrier density, this has become clear as a result of work on the quantum spin Hall effect [12]. A key feature of these models is the appearance of massless topological states which replace the usual Fermi surface. These topological states are characterized by the appearance of persistent chiral spin currents which violate parity symmetry. Actually the topological states predicted by Haldane have not been seen because spin-orbit interactions are weak in graphene. However, massless topological states very similar to those predicted by Haldane are expected in bismuth, which can be thought of as a stack of graphene-like layers, and where spin-orbit effects are known to be large [13]. The topological chiral edge states in Bi have not yet been seen directly, but massless surface states have been seen in BiSb alloys [14].

Our theory of massless topological states in a layered metal with low carrier density is based on the idea that spin-orbit interactions can give rise to local chiral spin currents that have exactly the same form as the spin Hall current in a semiconductor:

$$j_\beta^\alpha = \sigma_s \varepsilon_{\alpha\beta\gamma} E_\gamma, \tag{1}$$

where $\sigma_s$ is the spin Hall "conductivity" and $E_\gamma$ is the local electric field. In a doped semiconductor $\sigma_s$ will be proportional to the Fermi momentum; while for the materials we are concerned with in this paper – the actinides – $\sigma_s$ is close to the quantized Hall conductance. Of course, inside a good conductor there can be no equilibrium long-range electric fields due to screening. However, in a material with a low carrier density there can be local electric fields – which can be particularly strong if there are lattice deformations which break inversion symmetry - and hence localized spin currents might arise in such a material even in the absence of an external field. In order to describe the effect of spin-orbit interactions in a layered non-degenerate conductor with a weak screening we introduce an effective magnetic field seen by the charge carriers which, neglecting spatial variations in the electric field, is given by the Chern-Simons equation

$$B_{\text{eff}} = -\frac{e}{\kappa}\rho, \tag{2}$$

where $B_{\text{eff}}$ is the magnitude of an effective magnetic field whose direction is perpendicular to the layer, $\rho$ is the charge per unit area, and $1/|\kappa|$ is an inverse length. If electric charge screening is weak $1/|\kappa|$ measures the strength of an "axionic" spin-orbit coupling and $B_{\text{eff}}$ can be interpreted as the effective magnetic field seen by charge carriers due to their motion in the local electric field.

The wave function for a 2-dimensional quantum gas of particles interacting via gauge potentials will satisfy a non-linear Schrodinger equation of the form [15]:

$$i\hbar \frac{\partial \psi}{\partial t} = -\frac{1}{2m}D^2\psi + eA_0\psi - g|\psi|^2\psi, \tag{3}$$

where $D_\alpha = \partial_\alpha - i(e/\hbar c)A_\alpha$. The gauge fields $A_0$ and $A_\alpha$ are not the usual electromagnetic fields that satisfy Maxwell's equations, but instead are determined self-consistently from the Chern-Simons Eq's (1-2) with $\sigma_s = \kappa$. The nonlinear term with coefficient $g$ represents the effect of spin-orbit coupling. It was shown some time ago [15] that analytical exact zero energy solutions to Eq.s (1-3) can be found if one assumes that a self-duality or anti-self-duality condition is satisfied:

$$g = \pm e^2\hbar/mc\kappa. \tag{4}$$

The corresponding 2-dimensional ground state contains spin polarized charge carriers with vortex-like spin currents and two units of effective magnetic flux attached to every carrier. The two signs for $g$ correspond to solutions where the carrier spins point either up or down along an axis perpendicular to the 2-dimensional plane. These exact 2-dimensional solutions were generalized to a layered material in ref. [10]. In the limit when the number of layers is very large the exact solution corresponding to Eq. (4) has the form [10];

$$\Psi = f(w) \prod_{k>j}^{\infty} \left[\frac{R_{jk} + U_{jk}}{R_{jk} - U_{jk}}\right]^{1/2}, \tag{5}$$

where $R_{jk}^2 = U_{jk}^2 + 4(z_j - z_k)(\bar{z}_j - \bar{z}_k)$, $U_{jk} = u_j - u_k$ where $u_k$ is the height of the layer, and $z_j$ a coordinate within the jth layer. $f$ is an entire function of the $\{\bar{z}_i\}$ in the self-dual case and $\{z_i\}$ in the anti-self-dual case. The wave function (5) can be thought of as describing a gas of monopole-like sources of the effective field $B_{eff}$, located at positions $X_j \equiv (u_j, z_j)$. In ref [10] these monopole-like objects were given the name "*chirons*". Writing the product on the rhs of Eq. (5) as exp(S) defines an effective action for a chiron:

$$S = \frac{1}{2} \sum_j \ln \frac{R_j + u - u_j}{R_j - u + u_j} , \qquad (6)$$

where $R_j^2 = (u - u_j)^2 + 4(z - z_j)(\bar{z} - \bar{z}_j)$. The wave function (5) resembles in some respects Laughlin's wave function for the fractional quantum Hall effect; for example, moving the $z$ coordinate of a chiron around the position of a chiron in a different layer changes S by $i\pi$. However, in contrast with the fractional quantum Hall effect, there are two distinct degenerate ground states corresponding to the self-dual and anti-self-dual solutions for both the 2-dimensional and the layered versions of Eq. (5). Physically these two solutions correspond to having all the carrier spins be either up or down, and they can form a Kramers pair – which may be the key to understanding high $T_c$ superconductivity [16].

Recently it has been pointed out [16] that an effective interaction of the form (6) leads to a Kosterlitz-Thouless-like transition in a gas of self-dual and anti-self-dual chirons. This transition differs from the classical KT transition because the effective interaction between chirons is only logarithmic when the average separation between chirons is less than the interlayer spacing. This means that the predicted transition temperature will depend on the density of chirons. Using the known c/a ratio and carrier effective mass for the high $T_c$ superconductor LSCO the predicted transition temperature as a function of doping is in remarkably good agreement with what is observed in the under-doped region [17]. In the over-doped region screening becomes important and the effective interaction is no longer logarithmic; leading to a suppression of the KT-like transition. A similar kind of transition may be operating in the "115" Pu and Np layered superconductors.

Our thesis is that the chiron theory of high $T_c$ superconductivity [16] can also be used to explain the unusual behaviour of the elemental actinides in the vicinity of Np/Pu/Am. In particular, we believe that the strong spin-orbit interactions in these elements lead to the formation of chiron-like topological solitons. In $\alpha$-Pu, Np, and perhaps even U these chiron-like solitons form a lattice, producing a pattern of localized spin currents that resemble orbital ordering. These solitons become massless at the $\alpha/\delta$ transition, which would lead to a KT-like transition in the absence of a first order transition. The dependence of this transition on disorder and the strength of the spin-orbit coupling provides a nice explanation for the low temperature $Pu_{1-x}Ga_x$ phase diagram [18] and why the $\alpha/\delta$ transition can be induced by radioactivity [19]. Beyond the $\alpha/\delta$ transition the actinide ground state should share some features in common with the superconducting state in high $T_c$ materials. For example, we would expect that PuAm alloys should exhibit a form of superconductivity resembling high $T_c$ superconductivity. In addition, *f*-electron pairs should form bubbles nucleated by impurities or lattice defects. Evidence for the existence of these bubbles may have already appeared in observations of the effect that self-damage from $\alpha$-radioactivity has on the low temperature magnetization of Pu [20].

The *piece de resistance* of the chiron theory of actinides is the behaviour of the conductivity near to the $\alpha/\delta$ transition. Beyond the intriguing fact that the room temperature conductivity of $\alpha$-Pu is anisotropic, both $\alpha$ and $\delta$-Pu have large room temperature resistivities [21]; indeed their resistivities are similar to those of high $T_c$ materials at temperatures just above high $T_c$. It is simple to check that the room temperature resistivity of δ-Pu (110 μΩ-cm) corresponds to a sheet resistance that is nearly quantized;

$$\rho(\delta\text{-Pu}) \approx (\hbar/4e^2) c , \qquad (7)$$

where $c$ is the lattice constant. It is a generic feature of the superconducting-insulator transitions in thin film superconductors that near to the transition the longitudinal resistivity and Hall conductivity satisfy a duality relation [5]:

$$\rho_{xx}/(h/4e^2) \approx \sigma_{xy}/(4e^2/h); \qquad (8)$$

i. e. the sheet resistance and Hall conductance when measured in quantized units are nearly the same. One way this can happen is that both quantities are nearly quantized. Physically this comes about because right at the transition 2 kinds of quasi-particles appear – conduction electron pairs and vortices in the thin film superconductor case and $f$-electron pairs and chirons in the actinide case – that are nearly interchangeable. It follows that Eq. (8) is satisfied because a flow of quantized vortices (or chirons) generates a Josephson voltage proportional to current whose coefficient is approximately the quantum of conductance. At finite temperatures a metal with a resistivity satisfying Eq. (7) is a "Planckian metal"; i.e. the effective Drude relaxation time is given by

$$\tau \approx \hbar/kT , \qquad (9)$$

where $k$ is Boltzmann's constant. The term Planckian metal was coined by Jan Zaanen in the context of high $T_c$ superconductivity [22]. It is somewhat surprising that it has apparently never been noticed before that near to the α/δ transition Pu is also a Planckian metal. This situation is reminiscent of quantum chromodynamics (QCD), where the vacuum state is a monopole condensate and quarks are confined inside hadrons, but at high temperatures an exotic quantum fluid containing quarks and monopoles appears with a nearly quantized viscosity [23].

In summary, it seems quite compelling that the α/δ transition in Pu is analogous to the superconducting-insulating transition in disordered thin film Type II superconductors, with spin orbit interactions playing the role of an applied magnetic field. The "insulating" δ-phase corresponds to a ground state with a chiron condensate and confinement of $f$-electron pairs. From the point of view of theoretical physics it is very exciting that impurity stabilized δ-Pu may be a condensed matter realization of a "dual vacuum", where the ground state contains a monopole-like condensate. The concept of a dual vacuum was originally introduced by Mandelstam to explain quark confinement, and it has been a struggle lasting decades to try and find a detailed model for such a state.

**Acknowledgments**

The author is very grateful for conversations with M. Fluss, S. Hecker, S. McCall, K. Moore, P. Soderlind, J. Zaanen, and S-C. Zhang.
This work was performed under the auspices of the U.S. Department of Energy by Lawrence Livermore National Laboratory under Contract DE-AC52-07NA27344.

Table 1.

Correspondence between the actinides
and thin film superconductors in a magnetic field

| Actinides | 2D Superconductor |
|---|---|
| U | Meisner superconductor |
| Np/$\alpha$-Pu | Electron condensate, vortex lattice |
| $\delta$-Pu | Quantum critical point |
| $Pu_{1-x}Am_x$ | Electron glass, vortex condensate |
| Cm, Bk,… | Mott insulator |

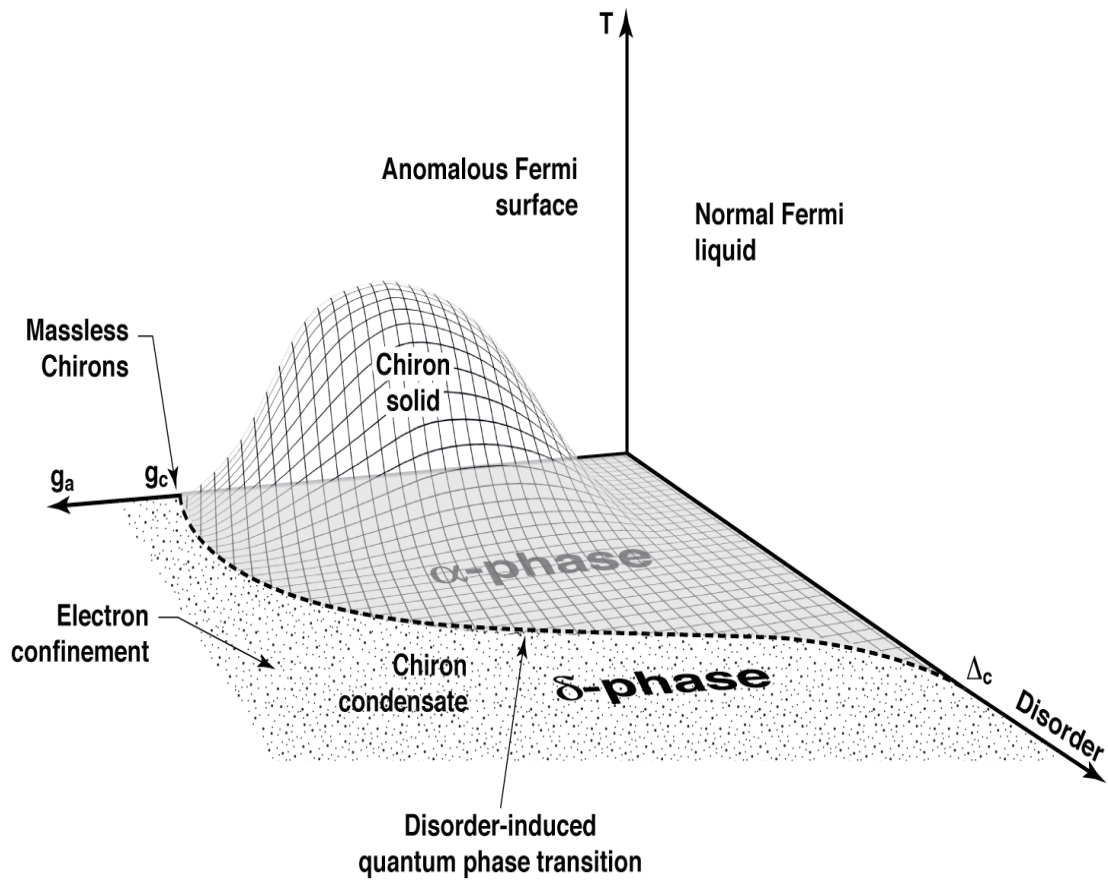

Fig. 1. Schematic phase diagram for a metal with low carrier density and a strong spin orbit coupling. At a critical value of the spin orbit coupling or disorder the soliton-like carriers condense to form a pair condensate. For weak spin orbit coupling or small disorder the low temperature phase contains a lattice of localized spin currents.